\documentclass[reprint,secnumarabic,amssymb,nobibnotes,aip,pra]{revtex4-2}

\usepackage{graphicx}
\usepackage[colorlinks]{hyperref}
\usepackage{overpic}
\usepackage{bm}
\usepackage{amsmath}

\newcommand{\ket}[1]{\vert #1 \rangle}

\makeatletter
\def\@email#1#2{%
 \endgroup
 \patchcmd{\titleblock@produce}
  {\frontmatter@RRAPformat}
  {\frontmatter@RRAPformat{\produce@RRAP{*#1\href{mailto:#2}{#2}}}\frontmatter@RRAPformat}
  {}{}
}%
\makeatother
\begin{document}

\preprint{AIP/123-QED}
\title{Modeling Line Broadening and Distortion Due to Spatially Non-Uniform Fields in Rydberg Electrometry }
\thanks{Publication of the U.S. government, not subject to U.S. copyright.}
\author{Andrew P. Rotunno}
\affiliation{National Institute of Standards and Technology, Boulder,~CO~80305, USA}
\author{Amy K. Robinson}%
\affiliation{Department of Electrical Engineering, University of Colorado, Boulder,~CO~80305, USA}
\author{Samuel Berweger}
\author{Nikunjkumar Prajapati}
\author{Alexandra B. Artusio-Glimpse}
\author{Matthew Simons}
\author{Christopher L. Holloway}
\affiliation{National Institute of Standards and Technology, Boulder,~CO~80305, USA}

\date{\today}

\begin{abstract}
We present a model for calculating broadened Autler-Townes spectra when probing ac electric field magnitudes by use of Rydberg atom electromagnetically-induced transparency. 
This model approximates the atom vapor as a multi-layered media and uses Beer's law to combine probe beam absorption through many discrete thin segments, replicating the broadening seen in experimental measurements.
The methodology can be easily applied to other non-uniform optical parameters when monitoring total spectral transmission or phase delay. 
Field non-uniformity is present in standing waves of unmatched waveguides and glass vapor cells, which are generally due to source characteristics and environmental reflections. 
We present broadening and distortion effects caused by various types of electric-field distributions, and compare this model to results obtained from measured atomic spectra.
\end{abstract}

\maketitle

\section{Introduction}
In recent years, Rydberg atom spectroscopy has been a fruitful method for making traceable measurements of radio-frequency (RF) electric (E) field amplitudes. 
Other electric and magnetic fields effects can also be observed via transmission spectroscopy\cite{tan1,sed1, holl1,9748947}.
In these sensors, Rydberg atom energies are observed via electromagnetically induced transparency (EIT). 
Spectral lines can be split linearly with applied resonant electric fields via the  Autler-Townes (AT) effect, and off-resonance, lines are shifted in energy by ac Stark effects. 
By measuring these effects carefully, we can measure amplitude \cite{gor1, sed1, 9748947, holl1, holl2, tan1, r5, gor2, gor3}, polarization \cite{sed2, access}, and phase \cite{sim3, jing1} of an RF field in a way that is traceable to International System of Units (SI)\cite{holl1} with various applications\cite{9748947}.

Any real-world transmission spectrum one measures will sample many atoms over some extended space, a long cylinder for the case of counter-propagating EIT. 
In general, we may expect that the RF field we measure is not uniform over this space, leading in general to broadening in the observed spectrum.
Non-uniformity in amplitude of the RF field can come from reflections, standing waves, spatial attenuation, and the method of generation, all of which can cause a varying field over the interrogation region\cite{r5, fan4, sahrai2007atom}. 
While one can minimize reflections using RF absorber foam and other particular experimental arrangements, we find that in almost any physical setup, including standard vapor cells, the observed line structures are generally broadened by a spatially non-uniform RF field, making it a nearly universal experimental issue for precision Rydberg electric field sensors.

\begin{figure}
    \centering
    \begin{overpic}[width=.48\textwidth]{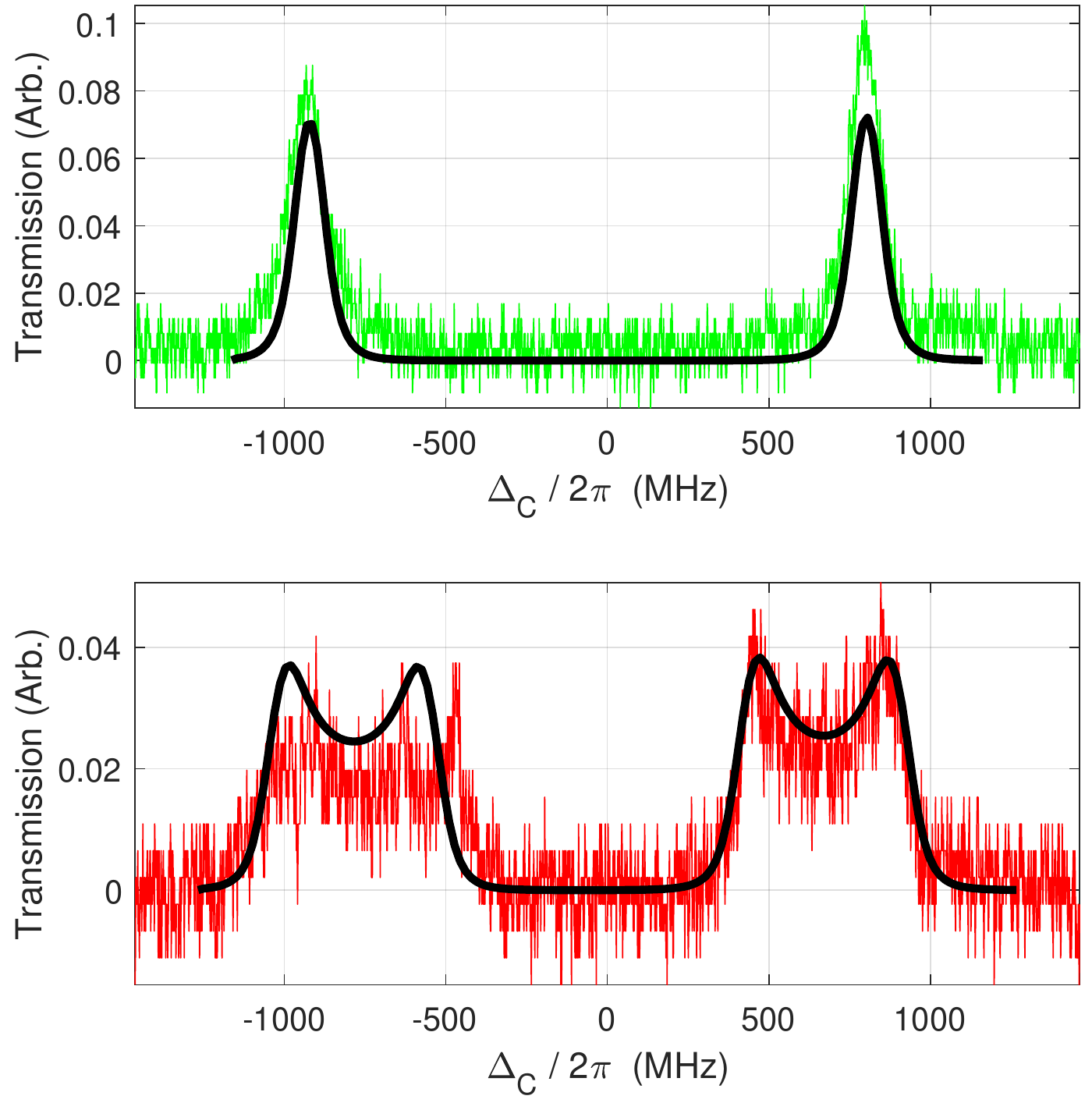}
    \put(90,94){(a)} \put(90,36){(b)} \end{overpic}
    \caption{Rydberg EIT in a waveguide due to (a) a constant RF electric field, and a (b) standing wave non-uniform field. 
    We compare spectrum data (color) to simulation curves (thick black). 
    Using values $\Omega_p/2\pi\approx80$~MHz, $\Omega_c/2\pi\approx2$~MHz, $\Delta_{RF}/2\pi=150$~MHz, $\Delta_p=0$, $\wp_{3,4}=1370~ea_0$, with electric fields (a) 98~V/m, and (b) a sine function spanning 55 to 110~V/m. }
    \label{fig:moneyshot}
\end{figure}

We present a model for calculating EIT spectra through spatially-varying electric field amplitudes longitudinally along the beam propagation path by discretizing the changing field and combining the local transmission values into an aggregate observed spectrum. 
This discretization method can generally be applied to any optical parameter.
In this paper, we consider spectral line broadening due to spatial non-uniformity in RF electric field amplitude as measured by the AT/EIT method. 
We demonstrate this in Fig.~\ref{fig:moneyshot}, comparing measured data with our model for atoms in a waveguide. 
When the electric field is a traveling wave, it is nearly constant over space, as in the measurement of a  well-matched waveguide mode Fig.~\ref{fig:moneyshot}(a).
However, when reflections induce a significant standing wave, as in a poorly matched waveguide, we observe the characteristic broadening shown in Fig.~\ref{fig:moneyshot}(b). 
Our work suggest that prior works where AT/EIT broadening was observed\cite{Fan2015, Robinson2021, gor2, robinson2021AoA,simons2018uncertainties, fan2014subwavelength, holloway2022electromagnetically, stack2016microwave, anderson2016optical, chopinaud2021optimal, simons2018fiber} might be reinterpreted in terms of spectral broadening caused by spatially non-uniform electric field amplitude.
Several groups have attempted to model the effect of electric field non-uniformities looking at the result of electric susceptibility $\chi$ for different electric field $|E|$ inputs; however, no resulting transmission curves were reported \cite{dubey20192, PhysRevA.94.013842, PhysRevA.94.013842, PhysRevA.84.043840}.
We believe this simulation method can help refine fitting for application-based field measurements with atoms.
That is, rather than losing precision due to field broadening, this technique can yield more information about the particular field's spatial curvature.

In Sec.~\ref{sec:calc}, we present the background on typical absorption calculations and our modification to account for a spatially varying field. 
In Sec.~\ref{sec:results}, we illustrate the principle with a number of example spectra calculated for various field arrangements using this method. 
In Sec.~\ref{sec:comparison}, we compare a standing-wave model to measured data observed in a waveguide, demonstrating the model's ability to characterize field distributions. 
We conclude in Sec.~\ref{sec:conc}.
In the appendix, we describe the master equation underlying the optical calculations in Sec.~\ref{sec:mastereq}, and we analyze multi-reflection effects from this layered model in Sec.~\ref{sec:level3}.

\section{\label{sec:calc} Spectrum Calculation and Modification}
The Rydberg resonant field measurement scheme has been used previously to give traceable measurements of electric field amplitudes.
The corresponding simulation of transmission for a four-level system is well-documented~\cite{tan1,sed1, holl1,9748947}.
Precision measurements using spectroscopy are often limited by the linewidths observed, which can become broadened by non-uniform energy shifts over the observation volume. 
Here, we focus on variation in the target measurement of RF electric field amplitude $|E(x)|$ along the laser path. 

\begin{figure}
    \centering
    \begin{overpic}[width=.48\textwidth]{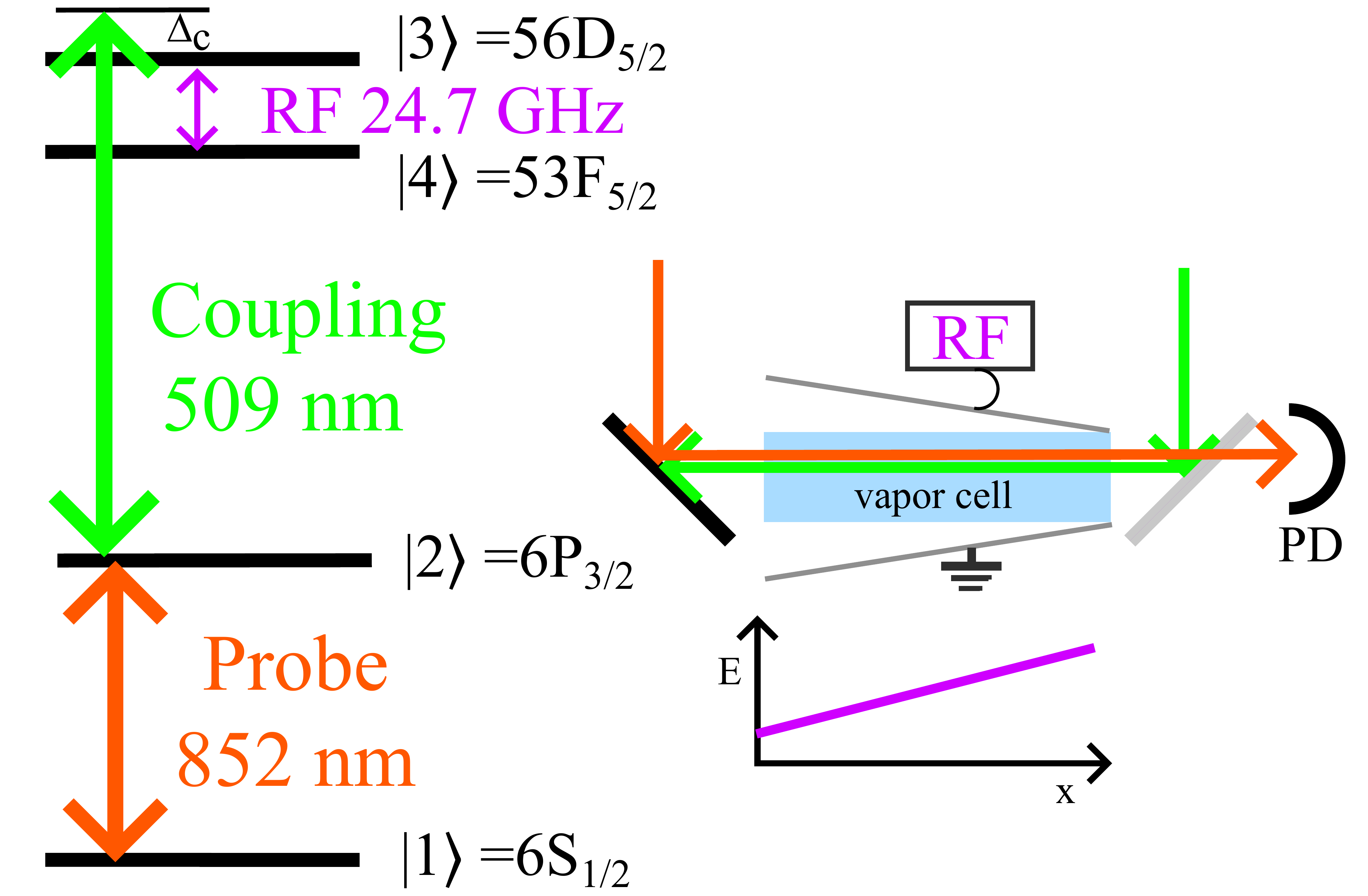}\put(70,60){(a)}\end{overpic}
    \\ \vspace{.2cm}
    \begin{overpic}[width=0.48\textwidth]{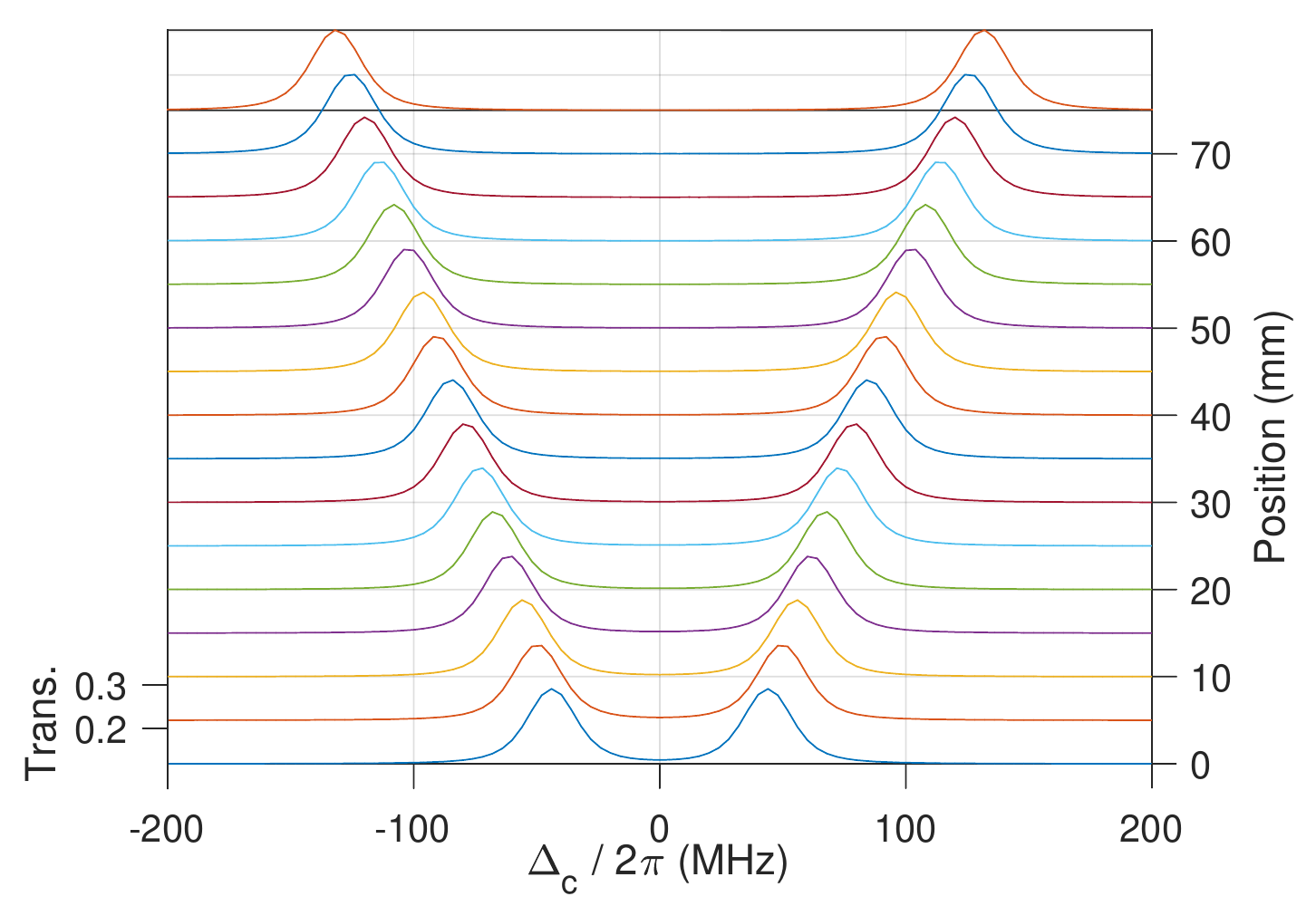} \put(5,60){(b)} \end{overpic}
    \caption{ Example of a linear gradient field measured by Rydberg EIT. 
    (a) Atomic level diagram in Cesium, and an example setup of non-parallel plates, which would produce a linearly changing non-uniform field, as sketched. 
    (b) Repeated spectrum calculations over space, linearly varying $E(x)$ from 5 to 15 V/m, sampling every 5~mm of a 75~mm cell.}
    \label{fig:discrete} \label{fig:diagram}
\end{figure}

The system we consider is illustrated in Fig.~\ref{fig:diagram}(a), consisting of two-photon EIT, where `probe' transmission on the D2 line ($|6S_{1/2}\rangle$ to $|6P_{3/2}$) through a Cesium vapor cell is monitored with a photodiode (PD), while the detuning $\Delta_c$ of a visible `coupling' laser is scanned across the resonance from the intermediate $P$ state to a highly-excited Rydberg state $|56D_{5/2}\rangle$, as illustrated in Fig.~\ref{fig:diagram}. 
When a radio frequency signal is resonant ($\Delta_{RF}=0$) to a strong Rydberg-Rydberg transition (to the  $|53F_{5/2}\rangle$ state) with large dipole moment $\wp_{3,4}$, the Rabi frequency $\Omega_{RF}$ gives the frequency-space `splitting' of the new observed line(s):
\begin{equation}\label{Rabi}
    \wp_{3,4}|E(x)| = \hbar \Omega_{RF}
\end{equation}
That is, the line splitting observed $\Omega_{RF}/2\pi$ is directly proportional to the field amplitude.

Using the model presented in Appendix \ref{sec:mastereq}, we calculate the expected spectrum using Beer's power absorption law of the probe laser with wavelength $\lambda_p$, over the extent of the vapor cell $L$, where the absorption is given by the imaginary part of the susceptibility $\chi$:
\begin{equation}\label{beers}
    T=\frac{P_{out}}{P_{in}} = \exp\left(- \frac{2 \pi }{\lambda_p} Im(\chi) L\right).
\end{equation}
As shown in Appendix \ref{sec:mastereq}, $\chi$ is related to the density matrix component ($\rho_{21}$) of the 4-level systems given in Fig.~\ref{fig:diagram}(a), where $\rho_{21}$ is obtained from the solution of the master equation.
We re-calculate $\rho_{21}(\Omega_{p,c,RF}, \Delta_{p,c,RF} )$ using selected parameters of the probe, coupling, and radio frequency electric field Rabi frequencies ($\Omega_p/2\pi\approx18$~MHz, $\Omega_c/2\pi\approx2.6$~MHz, and $\wp_{3,4}$ is nearly 17.5 MHz / (V/m) in these simulations, detunings ($\Delta_p=\Delta_{RF}=0$), and atomic constants.  Scanning over $\Delta_c$, we can string susceptibility calculation points together into transmission curves using Beer's law, Eq.~\ref{beers}.

In this single-medium Beer's law absorption model, all parameters are assumed fixed over the entire length $L$. 
When parameters are known to vary across the sample region, we require further analysis such as the one presented in this work to account for the spreading linewidths of transmission spectra.
Any imperfection in field flatness will grow with field amplitude, making large field measurements less precise due to the non-uniformity translated into wider spectral features with larger uncertainties. 
One may insert an ansatz broadening which scales with field into the master equation in the original single segment Beer's law model, but as we will demonstrate, this simple T2-style broadening would not be able to fully characterize the features observed due to non-uniform fields. 
There are further optical effects from medium non-uniformity that can be considered \cite{egan2012optical}. 
This work represents a first-order expansion to account for particular types of line broadening in transmission spectrum. 

To model the effect from a measurement over a non-uniform $|E(x)|$, we have discretized the space into $N_S$ segments, and calculated the transmission profile for a local $|E(x_i)|$ amplitude over each segment $dL = L/N_S$.
We generate a computation vector of electric field samples, whether invoked by function or sampling an EM solver solution. 
We show an example of this calculation in the spectra of Fig.~\ref{fig:discrete}(b) for a linear-over-space change in $|E(x)|$ from 5 to 15~V/m over 75~mm of cell length, which might be produced by un-parallel plates, shown in the schematic of Fig.~\ref{fig:diagram}(a).
Having generated $\chi$ across a scan of $\Delta_c$ for many samples of $E(x)$ or $\Omega_{RF}(x)$, one must combine them into an aggregate spectrum.

The absorption through each segment can be calculated using Beer's law for local susceptibility $\chi(x_i)$ over length $dL$. 
To combine transmission spectra, we can multiply sequential transmissions over each of the $N_S$ segments for each value of $\Delta_c$.
To simplify multiplying many exponentials, we can sum inside the exponential:
\begin{equation}
   T= \frac{P_{out}}{P_{in}} =  \exp\left(-\frac{2 \pi L}{\lambda_p N_S} \sum_{i=1}^{N_S} Im\left(\chi(x_i)\right) \right)
    \label{eqn_nLayer}
\end{equation}
where we have taken the constant $dL=L/N_S$ out of the sum. 
In essence, we `average' over $N_S$ different susceptibility curves, assuming each sampled field is held constant over $L$, as plotted in Fig~\ref{fig:discrete}. 
In this form, $\chi(x)$ can depend on any number of parameter changes, and gives total phase delay as well. 

The sampling density is important to converge to a smooth physical curve, since sparse sampling with large jumps in peak position from $\Omega_{RF}(x)$ to $\Omega_{RF}(x+dL)$ will leave `spikes' in the resultant curve. 
The sampling number $N_S$ can be raised to an arbitrary spatial resolution, which is useful for narrow linewidths typically used in precision experiments. 
In the simulations below, we use a larger probe power ($\Omega_p/2\pi\approx18$~MHz) than typical precision experiments($\Omega_p/2\pi\approx1$~MHz), using small values of $N_S$ to reduce the computation time for these examples. 

We will note that this broadening gives some additional information about field non-uniformity experienced by the atoms, giving many local measurements of the Rabi frequency \cite{sahrai2007atom}. 
These new peak features can be fit for parameters such as standing wave amplitude, field gradient, etc. by fitting observed line shapes to ansatz functions. 
These types of measurements might be especially useful as field monitors within other systems, such as our motivating case of low-loss power monitoring within a traditional waveguide. 

Beyond this approximation, using Beer's law over multiple segments, one can  consider probe attenuation, where we feed forward the loss in $\Omega_p$ over the $\Delta_c$ scan into the parameters of next segment.
We did not attempt this, as the segmented Beer's law approach was largely sufficient to reproduce the spectral features observed, although many further effects can be studied when precision transmission values are desired. 
Another effect to consider is outlined in Section~\ref{sec:level3}, where small reflections due to a spatially-varying index of refraction can interfere with other forward and reflected waves. 
This analysis is particularly useful in discontinuous cases such as the transverse waveguide measurement. 
However, we only present the simple Beer's law (Eq.~\ref{eqn_nLayer}) for the following curves. 


\begin{figure*}
    \centering
    \begin{overpic}[width=0.33\textwidth]{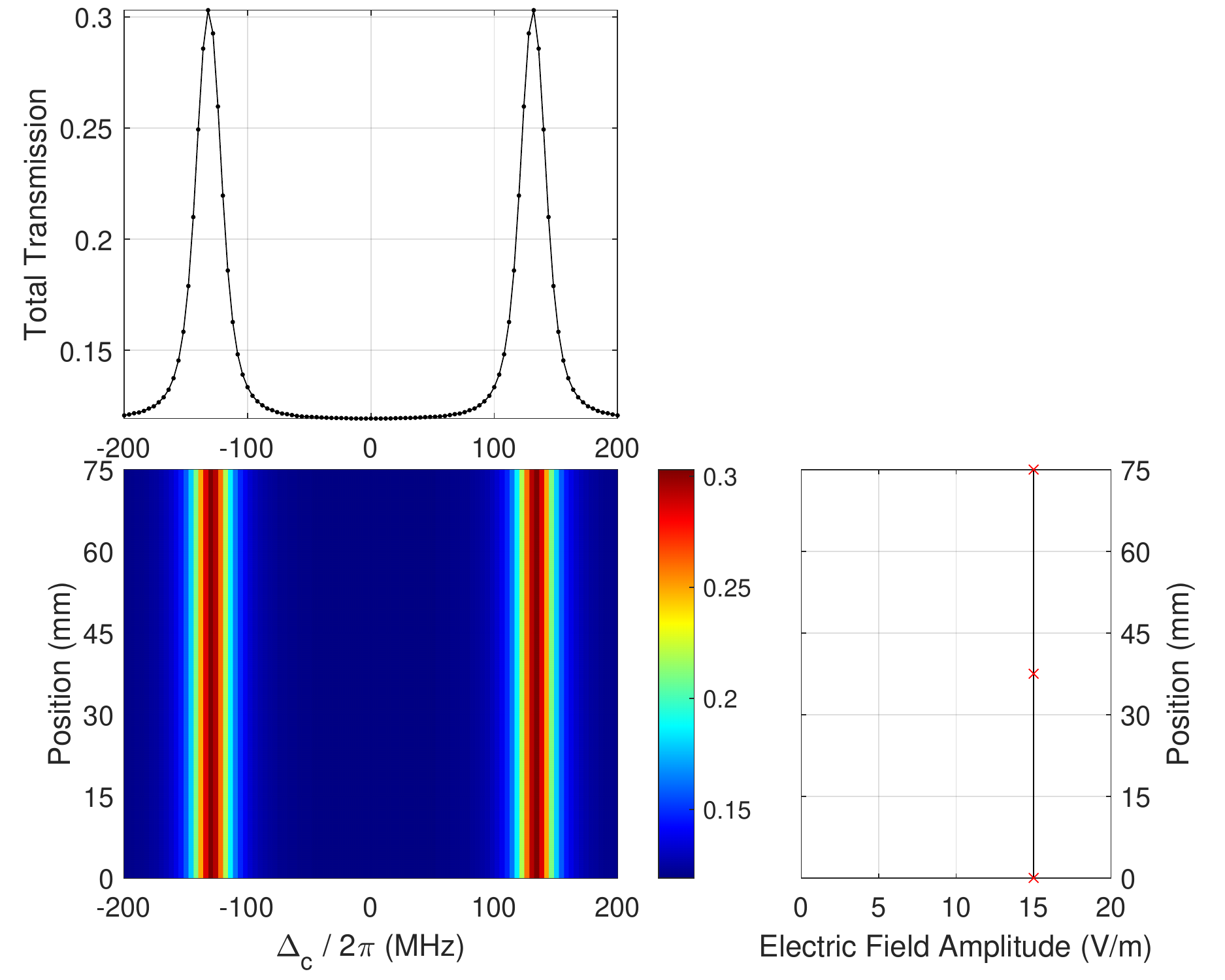}
    \put(58,70){\large (a)} \put(58,60)    {\large Constant} \end{overpic}\vrule
    \begin{overpic}[width=0.33\textwidth]{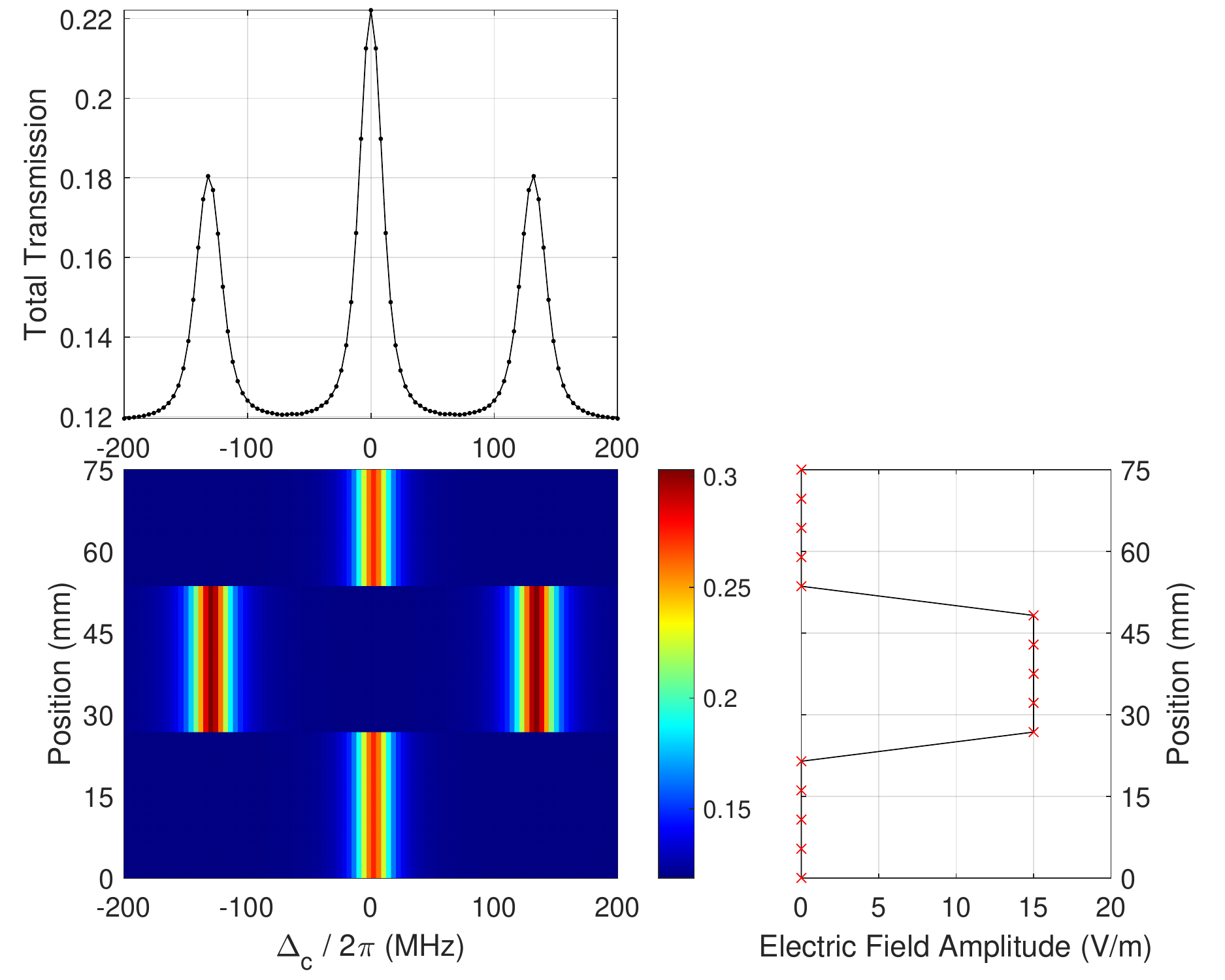}
    \put(58,70){\large (b)} \put(58,60)    {\large Step-wise} \end{overpic}\vrule
    \begin{overpic}[width=0.33\textwidth]{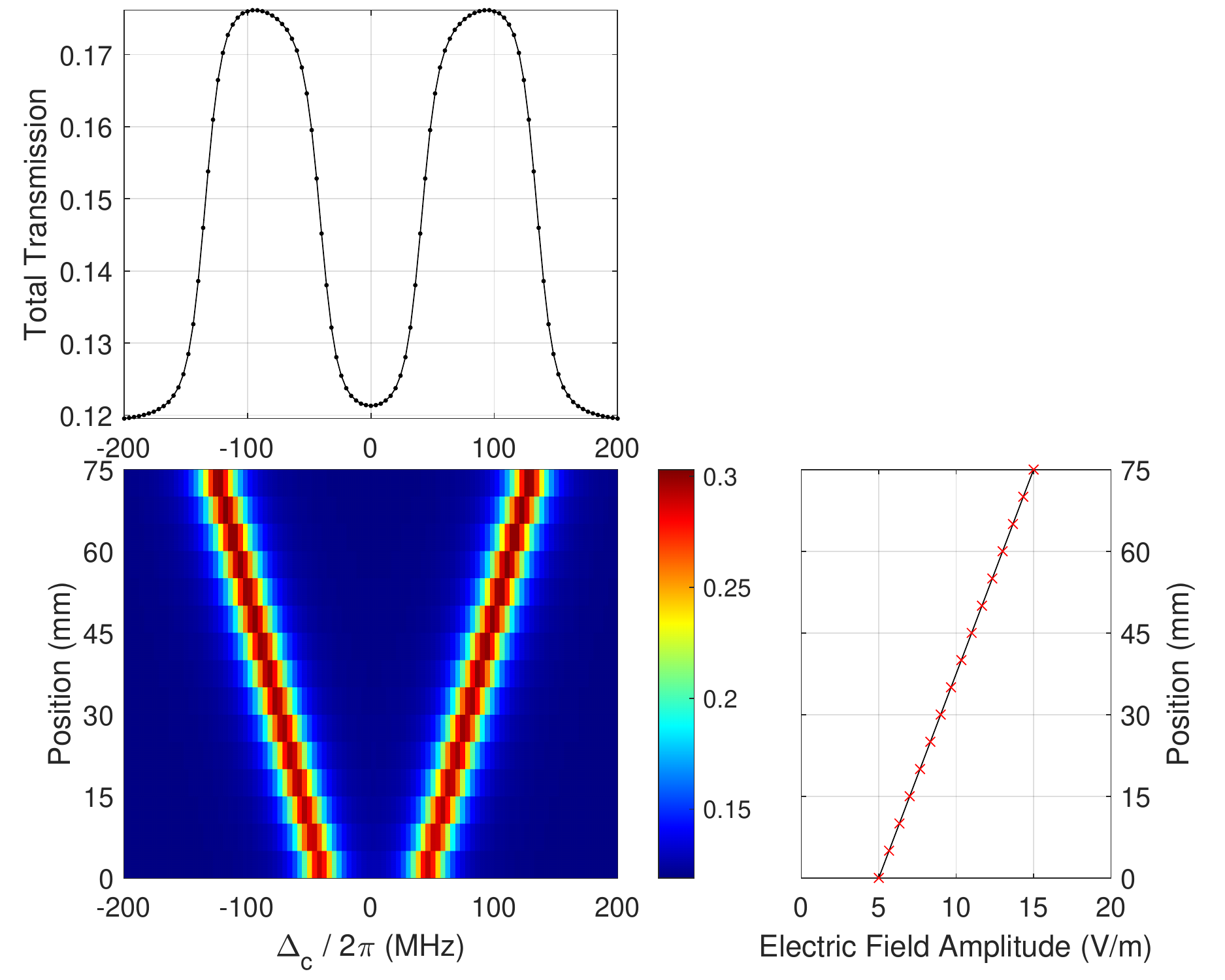}
    \put(58,70){\large (c)} \put(58,60)  {\large Linear} \end{overpic}\\
    \vspace{.25cm} \hrule \vspace{.25cm}
    \begin{overpic}[width=0.33\textwidth]{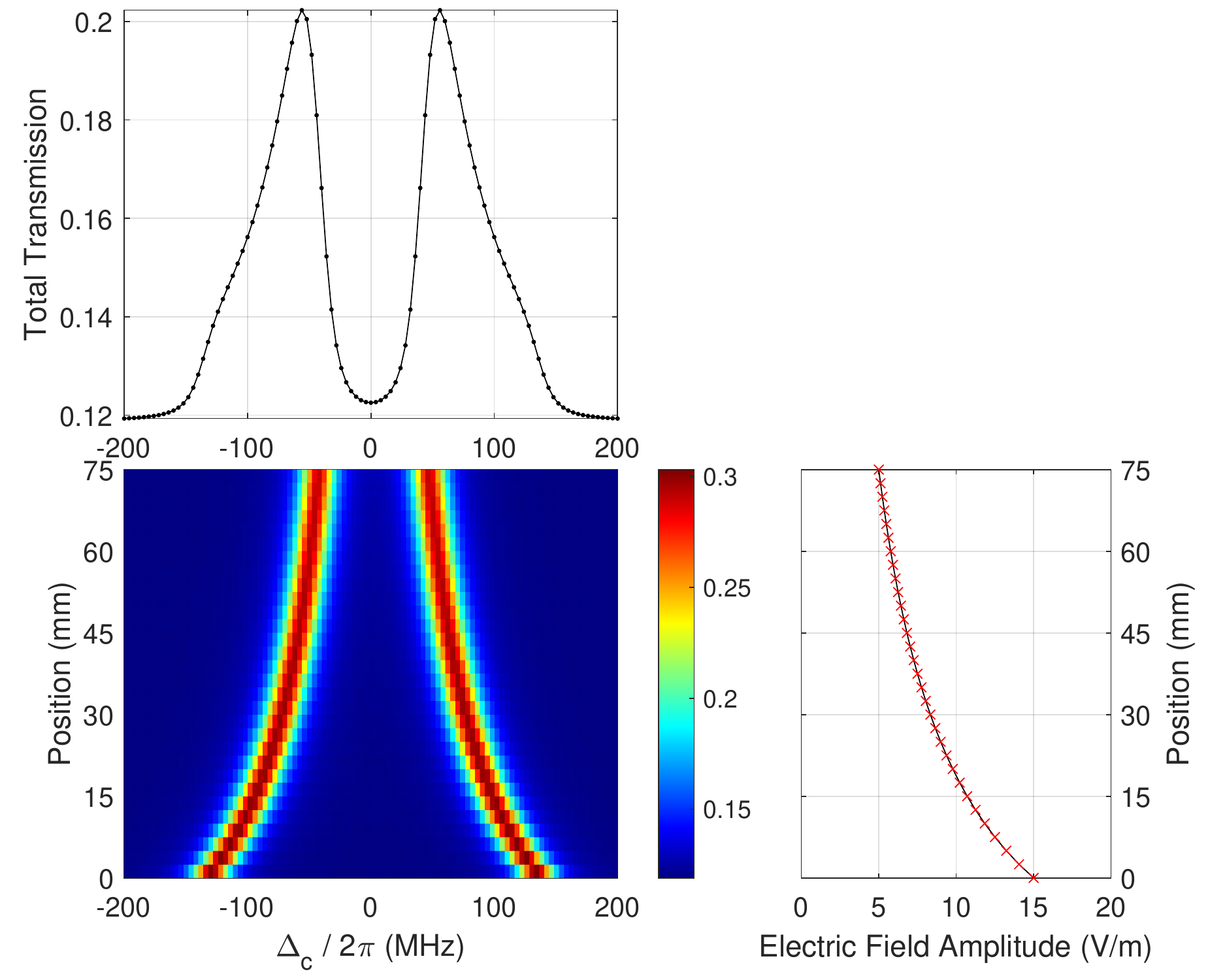}
    \put(58,70){\large (d)} \put(65,60)    {\large 1/$r$ } \end{overpic}\vrule
    \begin{overpic}[width=0.33\textwidth]{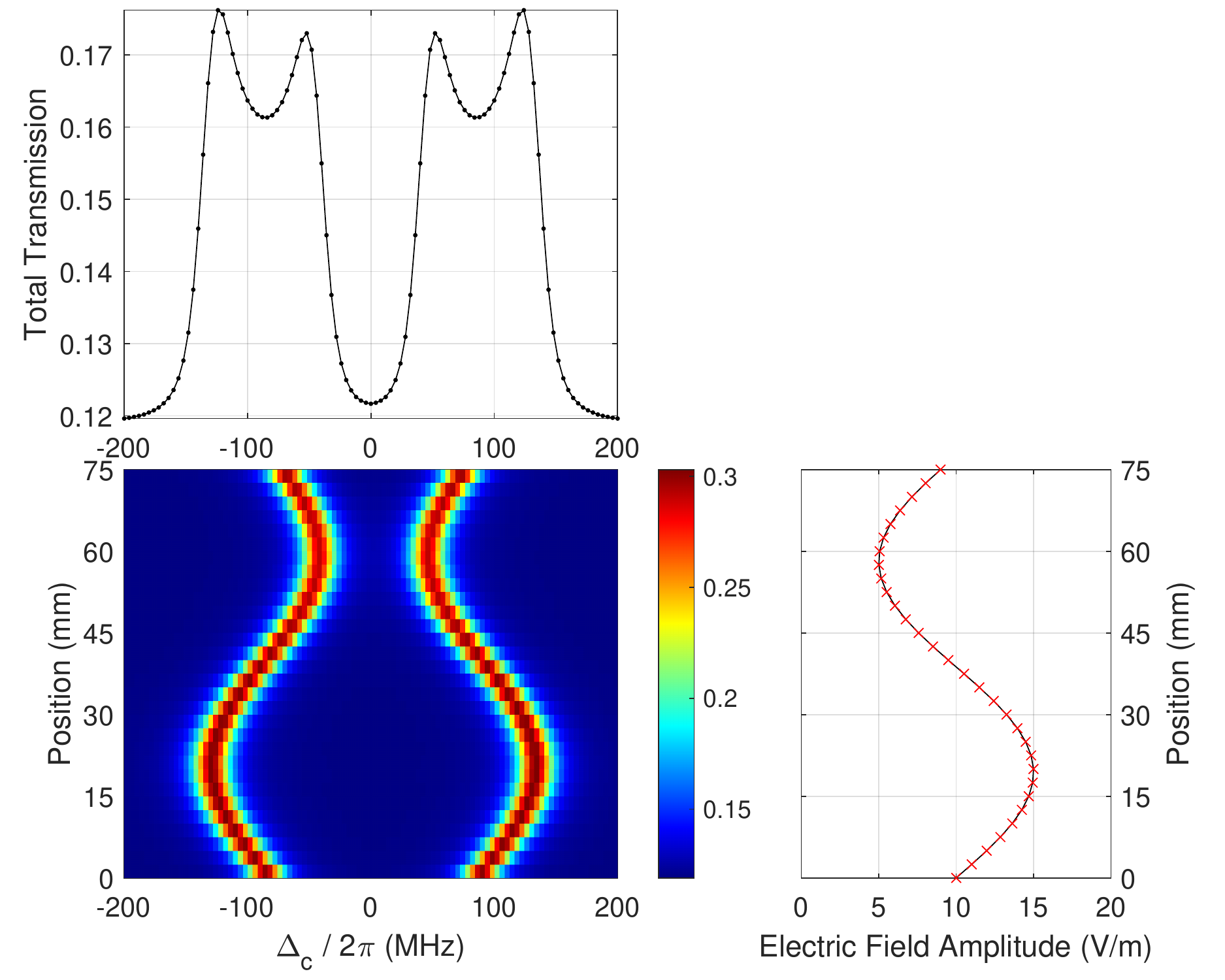}
    \put(58,70){\large (e)} \put(58,60)    { Standing Wave} \end{overpic}\vrule
    \begin{overpic}[width=0.33\textwidth]{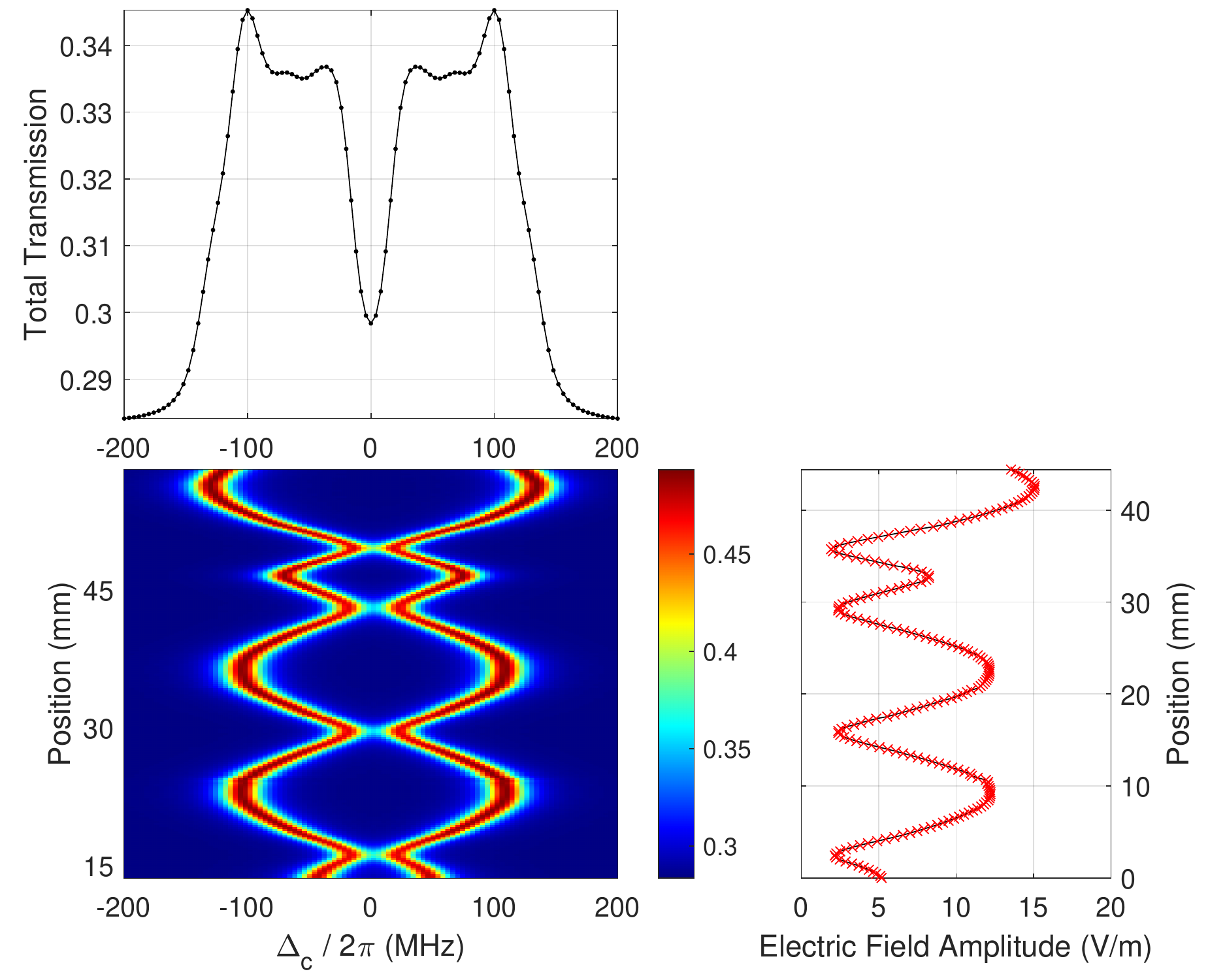}
    \put(58,70){\large (f)} \put(58,60) { Waveguide} \put(58,52)  { Simulation} \end{overpic}
    \caption{ Simulated total transmission profiles for a variety of cases (a-f) as labelled, and described in Sec.~\ref{sec:results}.
    For each case, we plot total transmission spectrum (top left), the local transmission spectra over space (bottom left), and the electric field over position (bottom right). 
    While $|E(x)|$ sampling is varied between parts, all simulations use $T=293$~K, $\Omega_p/2\pi\approx18$~MHz, $\Omega_c/2\pi\approx2.6$~MHz, $\Delta_p=\Delta_{RF}=0$, and $\wp_{3,4}/h\approx17.5$~MHz/(V/m) }
    \label{fig:fieldguide}
\end{figure*}
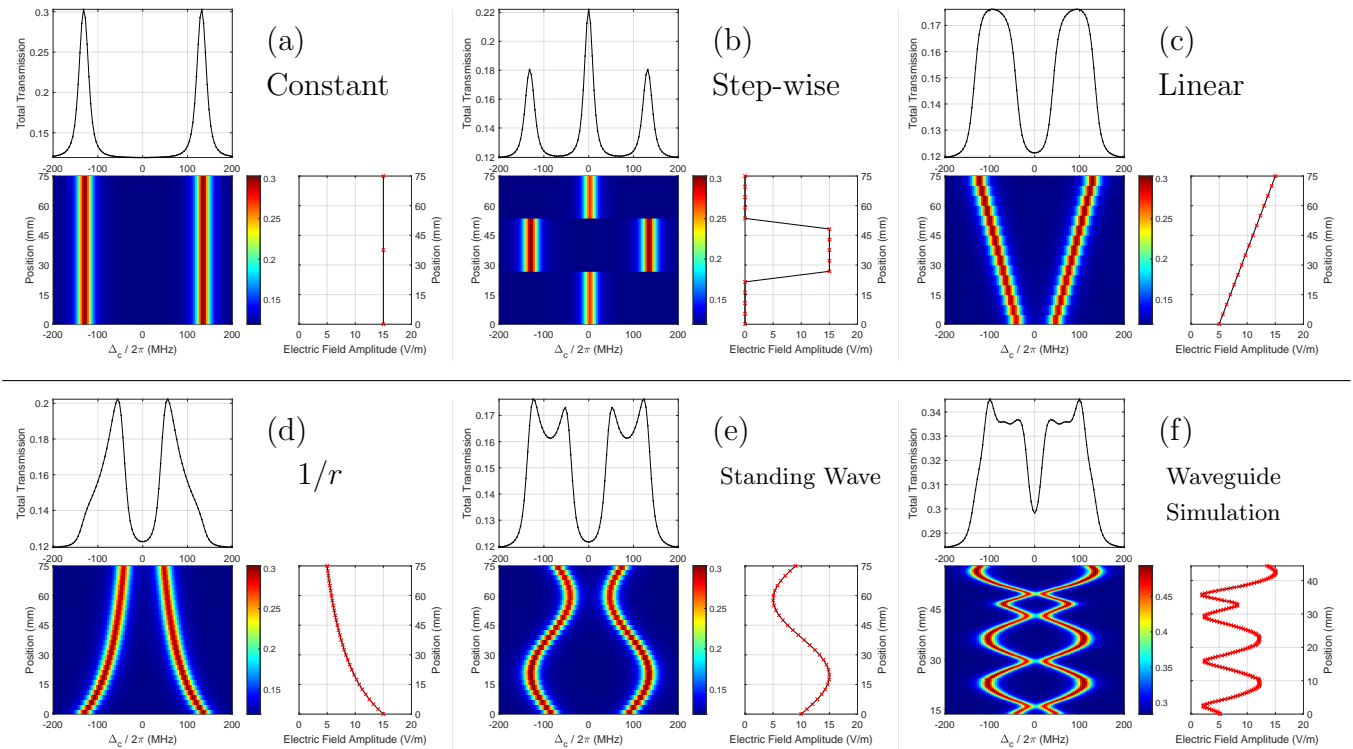

\section{Model Results\label{sec:results}}
Here, we provide a sample of some common field curvatures, and use our simulation method to illustrate the expected probe transmission curves using the method outlined in Sec.~\ref{sec:calc}.
Each sub-figure of Fig.~\ref{fig:fieldguide}(a-f) shows the total transmission plot (top left) aggregated over the entire cell length. 
The heatmap surface plot (bottom left) gives transmission spectrum over detuning horizontally and vertically over the position along the cell. 
Electric field along the cell is plotted as well (bottom right). 
Each field has a maximum of 15~V/m for direct comparison.

The constant case is shown in Fig.~\ref{fig:fieldguide}(a). 
This is synonymous with the single-segment Beer's law approach, and is used as a reference linewidth for the other lineshapes.

The step-wise case is shown in Fig.~\ref{fig:fieldguide}(b).
This case is physically realized in transverse waveguide probing, where atoms are observed in pinholes before and after the central waveguide, which has a constant field within it. 
We see a large unperturbed peak at the center from the section with no RF field, as well as the expected split AT peaks, roughly half the height of the center peak (as 1/3 and 2/3 of total $L$).

The simple gradient case is shown in Fig.~\ref{fig:fieldguide}(c), combining the curves of Fig.~\ref{fig:discrete}(b).
This linear field gradient represents a first order modification of line broadening, demonstrating the loss in peak transmission, as well as broadening that characterizes effects from field non-uniformity. 
For experimental sources, which cause a slight field gradient, the gradient usually grows in proportion to the field amplitude, causing linewidths to grow with applied power. 
This case illustrates that, to a first-order approximation of a gradient in the field, the total FWHM observed is approximately the sum of the total range of $\Omega_{RF}(x)$ induced by $|E(x)|$, plus the non-broadened EIT linewidth. 

The $1/r$ case is shown in Fig.~\ref{fig:fieldguide}(d).
This case represents the standard dipole antenna fall-off, or that from an RF horn, although they are typically employed transverse to the sampling path, not along it. 
Note the asymmetric peak shape is weighted to the lower end of the field range, which is proportional to the spatial extent of atoms exhibiting transmission at those frequencies. 

The sine or standing wave case is shown in Fig.~\ref{fig:fieldguide}(e).
This field motivated the present investigation, accounting for line broadening as we measure the electric field longitudinally in an un-matched waveguide. 
This sine field is used with different values for Fig.~\ref{fig:moneyshot}.
Note the outside `devil horn' characteristic, a result of the sine's sampling density near the extrema.

The case of a simulated standing wave is shown in Fig.~\ref{fig:fieldguide}(f). 
This field distribution comes from a computational EM solver software for the field inside of a waveguide with two glass slabs which cause the unmeasured poor matching, similar to the models used for Fig.~\ref{fig:longWG}. 
Significantly higher sampling density was required, owing to the spatial gradients involved. 
The simulation was performed near 20~GHz, but we treat the electric field value as if it were resonant.

\section{$|E(x)|$ Distributions: Simulated vs Measured\label{sec:comparison}}
In order to validate the application of this model to experimental results, we show a few different real-world examples compared to theoretical curves.
Details about the measured atom-filled waveguide can be found in author Robinson's thesis \cite{robinson2022microwave}. 

\begin{figure}
    \centering
    \includegraphics[width=.48\textwidth]{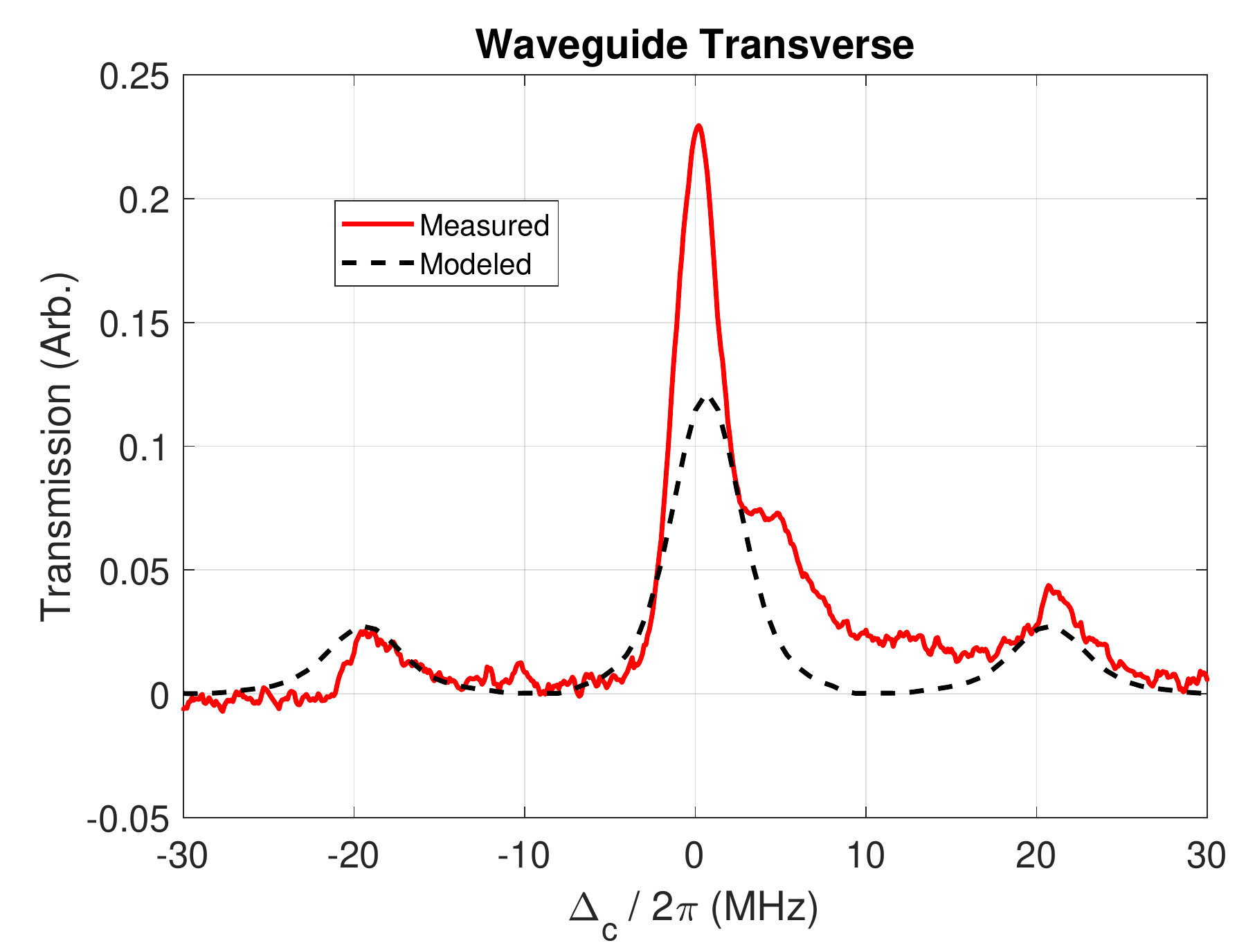}
    \caption{Comparison of data and simulation for a transverse measurement through a waveguide, which exhibits step-wise field behavior.}
    \label{fig:3layer}
\end{figure}

The first example in Fig.~\ref{fig:3layer} is a step-wise field that is sampled transversely across the center of a waveguide, where the field is nearly constant. 
The central waveguide portion contains atoms, but also two long optical access holes on either side, which do not contain the waveguide field. 
In this case, most of the atoms remain unaffected by the field, exhibiting the resonant EIT peak, while a portion of the atoms in the field are AT split as usual. 
There are additional asymmetric effects that distort the spectrum beyond the non-uniformity, including reflection EIT peaks which are difficult to eliminate in this narrow geometry.

\begin{figure}
    \centering
    \includegraphics[width=.48\textwidth]{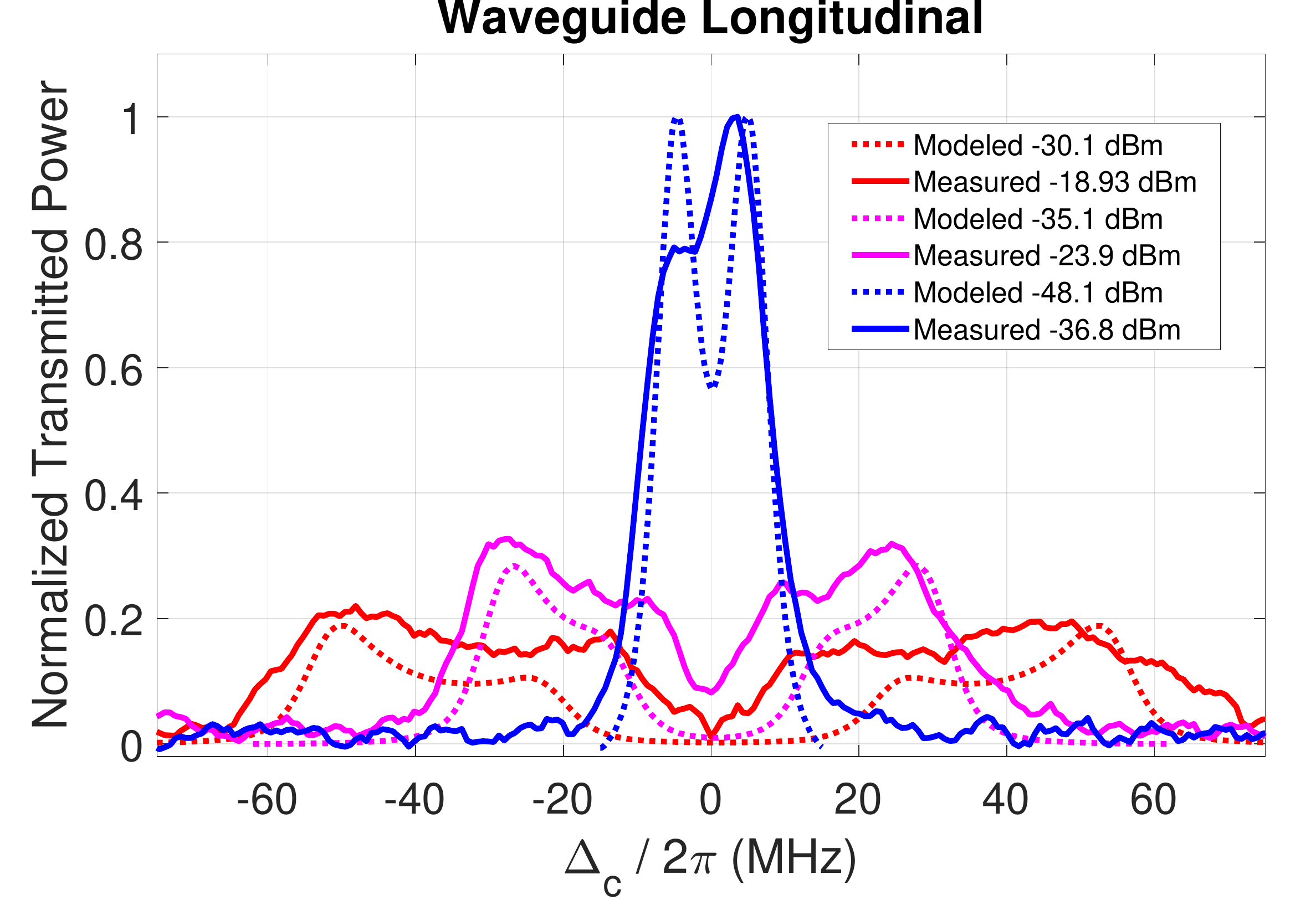}
    \caption{Comparison of data and simulation for a transverse measurement through a waveguide, which exhibits standing wave behavior. The field distribution in the waveguide is given by an EM solver. }
    \label{fig:longWG}
\end{figure}

A second example in Fig.~\ref{fig:longWG} considers a longitudinal standing wave along a waveguide.
Inserting glass into the waveguide structure to hold atoms also induces a standing wave in the travelling microwave signal, although its effects can be mitigated using two additional waveguide stub-like tuning elements.
We simulate the field inside of the waveguide using an EM solver for three different power values, each exhibiting varying levels of broadening. 
We note that a constant offset of $\approx11$~dB between measured and modeled curves accounts for significant insertion loss into the waveguide in the laboratory using directional couplers, whereas the model was fed by an appropriate waveguide mode.

The last example, shown in Fig.~\ref{fig:moneyshot}, illustrates an optical measurement longitudinally along waveguide with (a) proper and (b) poor tuning of stub-like cavities perturbing the waveguide to induce a standing wave. 
This theoretical curve uses a simple sine wave to approximate the lineshape, which largely captures the characteristic broadening without relying on an EM solver.

\section{Conclusion}\label{sec:conc}
We have presented a computational method to approximate broadening of Rydberg EIT lines due to spatially non-uniform electric field amplitudes. 
We have segmented the optical path, calculated local transmission values, and then combined these into a composite transmission spectrum. 
The calculation method is generic to any alteration in electric susceptibility parameters when total transmission or phase delay is monitored, and can therefore help bridge the gap between observed lineshapes and fitting theory curves. 
Rather than losing information from broadening, this method can enable additional information about spatial field variations to be fit from observed transmission spectra.

\appendix
\section{Master-equation model}\label{sec:mastereq}

We use a master-equation model of the EIT signals for the atomic transition schemes used here, shown in Fig.~\ref{fig:diagram}. 
The model presented here is for $^{133}$Cs atoms, although a similar model for $^{87}$Rb is presented in Ref. \cite{holloway2017}.  
The power of the probe beam measured on the detector (the EIT signal, i.e., the probe transmission through the vapor cell) is given by \cite{yarin}
\begin{equation}
P_{out}=P_{in} \exp\left(-\frac{2\pi L \,\,{\rm Im}\left[\chi\right]}{\lambda_p}\right)=P_{in} \exp\left(-\alpha L\right) \,\,\, ,
\label{intensity}
\end{equation}
where $P_{in}$ is the power of the probe beam at the input of the cell, $L$ is the length of the cell, $\lambda_p$ is the wavelength of the probe laser,  $\chi$ is the susceptibility of the medium seen by the probe laser, and $\alpha=2\pi{\rm Im}\left[\chi\right]/\lambda_p$ is Beer's absorption coefficient for the probe laser.  The susceptibility for the probe laser is related to the density matrix component ($\rho_{21}$)  by the following \cite{berman}
\begin{equation}
\chi=\frac{2\,{\cal{N}}_0\wp_{12}}{E_p\epsilon_0} \rho_{21_D} =\frac{2\,{\cal{N}}_0}{\epsilon_0\hbar}\frac{(d\, e\, a_0)^2}{\Omega_p} \rho_{21_D}\,\,\, ,
\label{chi1}
\end{equation}
where $d=2.02$ is the normalized transition-dipole moment \cite{SteckCsData} for the probe laser and $\Omega_p$ is the Rabi frequency for the probe laser in units of rad/s. 
The subscript $D$ on $\rho_{21}$ presents a Doppler averaged value. ${\cal{N}}_0$ is the total density of atoms in the cell and is given by
\begin{equation}
{\cal{N}}_0= \frac{p}{k_B T} \,\, ,
\label{nn}
\end{equation}
where $k_B$ is the Boltzmann constant, $T$ is temperature in Kelvin, and the pressure $p$ (in units of Pa) is given by \cite{SteckCsData}
\begin{equation}
p=10^{9.717-\frac{3999}{T}} 
\label{ppp}
\end{equation}
In eq. (\ref{chi1}), $\wp_{12}$ is the transition-dipole moment for the $\ket{1}$-$\ket{2}$ transition, $\epsilon_0$ is the vacuum permittivity, and $E_p$ is the amplitude of the probe laser E-field.

The density matrix component ($\rho_{21}$) is obtained from the master equation \cite{berman}
\begin{equation}
\dot{\boldsymbol{\rho}}=\frac{\partial \boldsymbol{\rho}}{\partial t}=-\frac{i}{\hbar}\left[\mathbf{H},\boldsymbol{\rho}\right]+\boldsymbol{\cal{L}} \,\,\, ,
\label{me}
\end{equation}
where $\mathbf{H}$ is the Hamiltonian of the atomic system under consideration and ${\boldsymbol{\cal{L}}}$ is the Lindblad operator that accounts for the decay processes in the atom. The $\mathbf{H}$ and $\boldsymbol{\cal{L}}$ matrices for the three different tuning schemes are given below.

We numerically solve these equations to find the steady-state solution for $\rho_{21}$ for various values of Rabi frequency ($\Omega_i$) and detunings ($\Delta_i$). This is done by forming a matrix with the system of equations for $\dot{\rho}_{ij}=0$. The null-space of the resulting system matrix is the steady-state solution.  The steady-state solution for $\rho_{21}$ is then Doppler averaged~\cite{berman}
\begin{equation}
\rho_{21_D}=\frac{1}{\sqrt{\pi}\,\, u}\int_{-3u}^{3u}\rho_{21}\left(\Delta'_p,\Delta'_c\right)\,\,e^{\frac{-v^2}{u^2}}\,\,dv\,\,\, ,
\label{doppler}
\end{equation}
where $u=\sqrt{2k_B T/m}$ and $m$ is the mass of the atom. We use the case where the probe and coupling laser are counter-propagating. Thus, the frequency seen by the atom moving toward the probe beam is upshifted by $2\pi v/\lambda_p$ (where $v$ is the velocity of the atoms), while the frequency of the coupling beam seen by the same atom is downshifted by $2\pi v/\lambda_c$.  The probe and coupling beam detuning is modified by the following
\begin{equation}
\Delta'_p=\Delta_p-\frac{2\pi}{\lambda_p}v \,\,\,{\rm and}\,\,\,
\Delta'_c=\Delta_c+\frac{2\pi}{\lambda_c}v \,\,\, .
\label{doppler2}
\end{equation}
We use the technique presented in Ref.~\cite{drew1} to decrease the computational time of the Doppler averaging procedure.

For the four level system, the Hamiltonian can be expressed as:
\begin{equation}
\begin{footnotesize}
H=\frac{\hbar}{2}\left[\begin{array}{cccc}
0 & \Omega_p & 0 & 0\\
\Omega_p & -2\Delta_p & \Omega_c & 0\\
0 & \Omega_c & -2(\Delta_p+\Delta_c) & \Omega_{RF}\\
0 & 0 & \Omega_{RF} & -2(\Delta_p+\Delta_c+\Delta_{RF})\\
\end{array}
\right]\,\, ,
\end{footnotesize}
\label{H4}
\end{equation}
where $\Delta_p$, $\Delta_c$, and $\Delta_{RF}$ are the detunings of the probe laser, coupling laser, and the RF source,
respectively; and $\Omega_p$, $\Omega_c$, and $\Omega_{RF}$ are the Rabi frequencies associated with the probe laser,
coupling laser, and the RF source. The detuning for each field is defined as
\begin{equation}
\Delta_{p,c,RF}=\omega_{p,c,RF}-\omega_{o_{p,c,RF}} \,\,\, ,
\label{rabi}
\end{equation}
where $\omega_{o_{p,c,RF}}$ are the on-resonance angular frequencies of transitions $\ket{1}$-$\ket{2}$, $\ket{2}$-$\ket{3}$, and $\ket{3}$-$\ket{4}$, respectively;
and $\omega_{p,c,RF}$ are the angular frequencies of the probe laser, coupling laser, and the RF source, respectively.
The Rabi frequencies are defined as $\Omega_{p, c, RF}=|E_{p, c, RF}|\frac{\wp_{p, c, RF}}{\hbar}$,
where $|E_{p, c, RF}|$ are the magnitudes of the E-field of the probe laser, the coupling laser, and the RF source, respectively.
Finally, $\wp_p$, $\wp_c$ and $\wp_{RF}$ are the atomic dipole moments corresponding to the probe, coupling, and RF transitions.

For the four-level system, the ${\cal{L}}$ matrix is given by
\begin{equation}
\begin{footnotesize}
{\cal{L}}=\left[\begin{array}{cccc}
\Gamma_2 \rho_{22} & -\gamma_{12}\rho_{12} & -\gamma_{13}\rho_{13} & -\gamma_{14}\rho_{14}\\
-\gamma_{21}\rho_{21} & \Gamma_3 \rho_{33}-\Gamma_2 \rho_{22} & -\gamma_{23}\rho_{23} & -\gamma_{24}\rho_{24}\\
-\gamma_{31}\rho_{31} & -\gamma_{32}\rho_{32} & \Gamma_4 \rho_{44}-\Gamma_3 \rho_{33} & -\gamma_{34}\rho_{34}\\
-\gamma_{41}\rho_{41} & -\gamma_{42}\rho_{42} & -\gamma_{43}\rho_{43} &  -\Gamma_4 \rho_{44}\\
\end{array}
\right] \,\,\, ,
\end{footnotesize}
\label{H4}
\end{equation}
where $\gamma_{ij}=(\Gamma_i+\Gamma_j)/2$ and $\Gamma_{i, j}$ are the transition decay rates. Since the purpose of this study is to explore the intrinsic limitations of Rydberg-EIT field sensing in vapor cells, no collision terms or dephasing terms
are added. While Rydberg-atom collisions, Penning ionization, and ion electric fields can, in principle, cause dephasing, such effects can, for instance, be alleviated by reducing
the beam intensities, lowering the vapor pressure, or limiting the atom-field interaction time. In this analysis we set $\Gamma_1~=~0$,
$\Gamma_2=2\pi\times$(6~MHz), $\Gamma_3=2\pi\times$(3~kHz), and $\Gamma_4=2\pi\times$(2~kHz). $\Gamma_{2}$ is for the D2 line in $^{133}$Cs \cite{SteckCsData}, and $\Gamma_{3}$ and $\Gamma_{4}$
are typical Rydberg decay rates.

\section{\label{sec:level3}Multi-layered Reflection Theory}
This section, we analyze the implications of multi-layer reflections, along the lines of traditional multi-layer media plane wave analysis. 
Reflection effects are physically significant for sharp changes in optical properties, such as the step-wise case presented above, and is negligible when susceptibility slowly varies over space. 

\begin{figure}
\centering
\includegraphics[width=.48\textwidth]{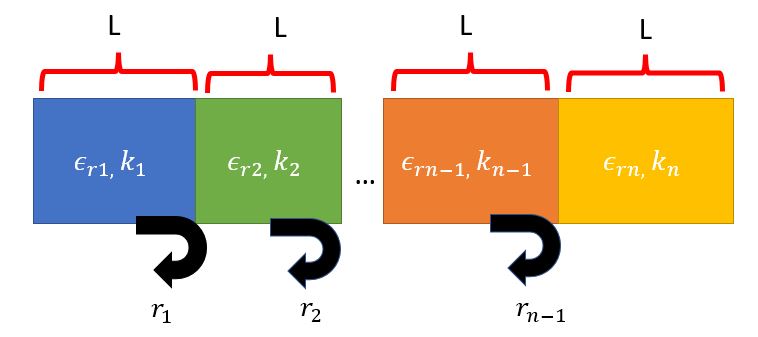}
\caption{The probe sees a multi-layered media that includes reflections due to changing susceptibilities of the dielectric slabs}
\label{refl}
\end{figure}
The first reflection coefficient, $r_{n-1}$, off of a two layer medium of relative permittivities $\epsilon_{r n-1}$ and $\epsilon_{r n}$, respectively is:
\begin{equation}
r_{n-1} = \frac{1-\frac{\sqrt{\epsilon_{r n}}}{\sqrt{\epsilon_{r n-1}}}}{1+\frac{\sqrt{\epsilon_{r n}}}{\sqrt{\epsilon_{r n-1}}}}
\end{equation} 
The reflection coefficient found at the beginning of slab $n-1$ is the reflection coefficient propagated across the medium of slab $n-1$
\begin{equation}
R_{n-1} = \frac{\sqrt{\epsilon_{r n}}+r_{n-1}\cdot e^{2k_{n-1}L}}{\sqrt{\epsilon_{r n}}-r_{n-1}\cdot e^{-2k_{n-1}L}}
\end{equation}
In this case, $k_{n-1} = \sqrt{\epsilon_{r n-1}} k = \sqrt{\epsilon_{r n-1}} \frac{2\pi}{\lambda_p}$ and $\epsilon_r$ is calculated from the susceptibility as $\epsilon_r = \chi+1$.
This reflection travels to the next layer in the medium, $n-2$. The total fields at the reflection point include both the reflected field from slab $n-1$ and the reflection coefficient from the incident field hitting the boundary of slab $n-2$ and $n-1$. 
\begin{equation}
r_{n-2} = \frac{R_{n-1}-\frac{\sqrt{\epsilon_{r n-1}}}{\sqrt{\epsilon_{r n-2}}}}{R_{n-1}+\frac{\sqrt{\epsilon_{r n-1}}}{\sqrt{\epsilon_{r n-2}}}}
\end{equation} 
The calculation loop repeats until the reflection reaches slab 1. Now that the reflection coefficients of the boundary are calculated, the total fields can be solved for using the standard free space field propagation equation:
\begin{equation}
E_1 = A_1e^{k_1L}+  B_1e^{k_1L}
\end{equation}
The value of $A_1$ is the electric field amplitude of the incident field in medium, and $B_1$ is the total reflected field in that medium slab. With the knowledge that the reflection coefficient of a slab is $r_1 = \frac{B_1}{A_1}$, this equation can be rewritten as
\begin{equation}
E_1 =A_1(e^{k_1L}+  r_1e^{k_1L})
\end{equation}
Due to continuity of fields, the incident field in slab 2, $A_2$ must equal $E_1$, therefore, $A_2$ is solved in terms of $A_1$
\begin{equation}
A_2 = \frac{A_1(1+r_1)}{e^{k_1L}(1+r_2e^{-2k_1L})}
\end{equation}

This loop continues until, finally, the incident field on the last slab $n$ is solved by 
\begin{equation}
A_n = A_{n-1}(1-r_{n-1})
\end{equation}
The final $A_n$ is the Fresnel transmission coefficient of the field, a ratio of the incident field and the output field. The total transmitted power will be $A_n^2*P_{out}$. In this case, $P_{out}$ is the transmitted signal caused by the standing wave and calculated in Eqn.~\ref{eqn_nLayer}. 

\begin{acknowledgments}
\noindent This work was partially funded by the NIST-on-a-Chip (NOAC) Program.
\end{acknowledgments}

\section*{AUTHOR DECLARATIONS}
\vspace{-3mm}
{\bf Conflict of Interest}\\
The authors have no conflicts to disclose.

\section*{Data Availability Statement}
Data is available at 
\href{https://doi.org/10.18434/mds2-2700}{doi:10.18434/mds2-2700}
\newpage

\bibliography{citations}
\vfill

\end{document}